# On the Compatibility Between Quantum and Relativistic Effects in an Electromagnetic Bridge Theory

*Massimo AUCI*
*Space Science Department*
*Odisseo Space*
*Via Battistotti Sassi 13, 20133 Milano, Italy*

The Dipolar Electromagnetic Source (DEMS) model, based on the Poynting Vector Conjecture, applied to dynamics of the electromagnetic transition ($p, \bar{p}$) $\rightarrow h\nu$, conduces in Bridge Theory to a derivation of the Lorentz transformation connecting pairs of events. The results prove a full compatibility between quantum and relativistic effects.

## 1) Introduction

Using the Bridge Theory (BT) [1] obtained from the Dipolar Electromagnetic Source (DEMS) model [2-3-4], we will prove, as the quantum effects described in BT are fundamental to justify self-consistently in the same context relativity and quantum phenomenology. Therefore, both behaviours are a consequence of a common electromagnetic origin (see ref. [2-3-4]) with a full formal and conceptual compatibility.

In the DEMS model, the transversal component of the Poynting vector is proved able to localise in space-time an amount of energy and momentum in agreement with quantum predictions for an exchanged virtual photon. BT explaining quantum phenomenology, allows us to formulate three fundamental statements:

I) when an elementary charged particle crosses the electromagnetic field of matter, in its action zone it interacts with all the other anti-charged particles sharing its energy and momentum producing a space-time distribution of DEMSs;

II) the theoretical value of the action constant that characterize the DEMS-model (see [1] and [4]) agrees with the Planck's constant of which it has the same role and value but not its fundamentality, since it depends on the internal structure of the electromagnetic field of the DEMS;

III) In the DEMS-model, quanta are self-consistent objects described by the interactions between charged particles pairs: *a DEMS is structurally and quantitavely equivalent to a photon having energy $E_s = \dfrac{hc}{\lambda}$ and momentum $\mathbf{P}_s = \dfrac{h}{\lambda}\hat{\mathbf{k}}$. The wavelength $\lambda$ is the minimum interaction distance achieved by a charged pair during their interaction.*

Following these three statements, if the particles of the pair move each other, their mutual interaction origins a DEMS with an energy and momentum settled by the minimal interaction distance $\lambda$. The amounts of energy and momentum agree with those of a photon of equal wavelength. In this sense, according to conservation laws, the DEMS converts the energy and momentum, associated to the relative motion of the two interacting particles, in a *virtual exchanged photon*.

We will consider the motion of the centre-of-mass (C.M.) of the DEMS towards an inertial observer: to except the observer placed in C.M., in each other frame an observer measures respect to the own line of view two different radial and transversal components of the total momentum **P** involved in the particles interaction. The components ($\mathbf{P}_\gamma, \mathbf{\Delta}$) are respectively the momentum of the photon emitted along the line of view of the observer and the momentum associated to the transversal motion of the C.M. of the DEMS respect the same line of view. At the same time selecting another observer, also the components change, i.e. for each inertial observer the wavelength measured is not the same.

## 2) Energy equation of a DEMS

To observe a charged particle we need to interact electromagnetically with it, producing a measurable effect. In general, when a charged particle as an elementary particle or an ion moves through matter, according to the statement (I) it interacts with all the other anti-charges inside its electromagnetic action zone. This effect produces a distribution of virtual photons, i.e. of active DEMSs along the path of the impinging particle. From the energetic point of view (statement (III)), the multi-interaction shares the energy and momentum of the impinging particle in a virtual space-time distribution of photons, which evolves in an overlapping of electromagnetic waves. To simplify this scenario, we will examine the production of just one DEMS.

In the model [1], the value of the elementary charges[1] is relevant only in order to obtain the exact value of the Planck's constant but not to describe the physical process.

The building process of the DEMS could be broken up into two different space-time phases that we name: A (alpha) and Ω (omega), corresponding respectively to the *born* and *death* phases of the DEMS:

(A-*alpha*): describes the approach between the interacting charges corresponding to the building of the DEMS. During this phase energy and momentum grow inside the source zone (see ref. [1]-[3]): in agreement with statement (III) the energy and momentum are equal to those of the exchanged photon;

(Ω-*omega*): describes the destruction of the DEMS. The impinging charged particles after the collision spring out of the source zone going away from one another with different energies and momenta; the wave emitted by the DEMS reduces the total amounts of energy and momentum driving the propagation of the out coming charged particles.

The source emits during both A and Ω phases, but only during the phase A a growing part of energy and momentum of the incoming particles is localised inside the source zone. In this phase, the DEMS production is equivalent to a photon exchange between the two particles.

When the Ω phase starts, only frames imbedded in the EM field of the DEMS can observe space-time source evolution.

An observer placed in an inertial frame *S* external to the C.M. of the DEMS, sees the energy and momentum carried by the interacting particles only in part transferred to the DEMS. In fact, because the relative motion between the C.M. of the source and frame *S*, only a part of the total energy and momentum involved supply respect the line of view of the observer the transversal motion of the C.M, the difference supply the DEMS.

---

[1] We assume the value of the elementary charge to be equal to one of the electron or positron.

In order to characterise dynamically the interacting particles with respect to an ideal observer at rest, we define their energies and momenta as $(E_1, \mathbf{P}_1)$, $(E_2, \mathbf{P}_2)$, so before the start of the phase A, the available total energy and momentum are

$$E = E_1 + E_2 \tag{1}$$

$$\mathbf{P} = \mathbf{P}_1 + \mathbf{P}_2 \tag{2}$$

where the momentum of each of the two interacting particles signed by the index $i = 1, 2$ can be written as $\mathbf{P}_i = m_i \boldsymbol{\beta}_i c$.

Considering the interaction occurring with energy and momentum (1)-(2) with respect to an observer $S$; let $E_\gamma$ and $\mathbf{P}_\gamma$ be energy and momentum of the observed photon, the energy and momentum conservation laws require:

$$\begin{cases} E_\gamma = P_\gamma c \\ \Delta_0^2 = (E - E_\gamma)^2 \\ \dfrac{\boldsymbol{\Delta} \cdot \boldsymbol{\Delta}}{c^2} = |\mathbf{P} - \mathbf{P}_\gamma|^2 \end{cases} \tag{3}$$

where $\Delta_0$ and $\boldsymbol{\Delta}/c$ are respectively the residual energy and momentum associated to the transversal motion of the DEMS respect to the line of view of the observer in the frame $S$.

Solving (3) respect the first equation we can write:

$$2E_\gamma(E - Pc \cos \varphi) - (E^2 - P^2 c^2) = \Delta^2 - \Delta_0^2. \tag{4}$$

where $\varphi$ is the angle between the total momenta $\mathbf{P}$ and the momentum $\mathbf{P}_\gamma$ associated to the photon observed along the direction of the line of view of the observer in the frame $S$.

In order to evaluate the right side $\Delta^2 - \Delta_0^2$ of the eq. (4), we consider the case of an observer placed in the C.M. of a DEMS produced during a head-to-head interaction. In this case, energy (1) and momentum (2) of the particles are seen completely involved in the DEMS formation, so residual energy $\Delta_0$ and transversal momentum $\Delta/c$ are identically equal to zero. This condition has as consequence that the right side squared difference in the equation (4) must be zero for each other observer imbedded in the same electromagnetic wave of the DEMS:

$$2E_\gamma(E - Pc \cos \varphi) - (E^2 - P^2 c^2) = 0. \tag{5}$$

Eq. (5) assigns to each inertial observer, with different line of view and relative motion respect the C.M. of the DEMS, different measured of photon energy, in fact solving eq. (5) with respect to $E_\gamma$ we obtain:

$$E_\gamma = \frac{E^2 - P^2 c^2}{2(E - Pc \cos \varphi)}. \tag{6}$$

## 3) Energy of a moving particle respect to an interacting rest frame

Considering an observer placed in the rest frame $S_i$ associated to one of the two interacting particles ($i$ = 1, 2), they have symmetrically the role of impinging particle and target. Let $\theta$ be the angle between the directions of the dipole moment $\mathbf{P}_s$ of the DEMS along the dipole axis and the momentum $\mathbf{P}$ of the impinging particle, according to the θ and φ angles definitions (see previous paragraph), we can rewrite eqs. (5) and (6) with $\varphi = \theta$.

If we place the observer at rest in the frame $S_2$ coinciding with the target particle 2, classically no energy and momentum are associated to this the observer.

From eqs. (1) and (2) we get

$$P \equiv |\mathbf{P}_1| = m\beta c, \ E = E_1, \qquad (7)$$

where $m = m_1$ and $\beta \equiv \beta_1$ are respectively mass and dimensionless velocity of the impinging particle 1.

If $\mathbf{R}$ is the vector connecting the two interacting particles, the impinging one is moving along the trajectory $r(t)$, interacting at a certain time $t$ at the physical conditions of time $t' < t$: usually the charge position $r' = r(t')$ at time $t'$ is delayed by a time $\Delta t = t - t'$ necessary for the signal to propagate along $\mathbf{R'} = \mathbf{R}(t')$ (see figure 1).

Let us call $R' = |\mathbf{R'}|$ the "effective" distance of interaction, and $R = |\mathbf{R}|$ the "actual" distance between the charges. If we assume that the particles achieve the minimal distance $R(0) = \lambda$ when $t = 0$, the interaction is delayed. Then the "effective" position $R'$ depends both on speed and angle of incidence at a previous time $t' < 0$.

Using figure 1, we can get an estimate of $R'$ (see also [1]) by writing

$$R' \leq \frac{v}{c} R' \cos\theta + R \ ,$$

that for $\beta = \dfrac{v}{c}$ yields

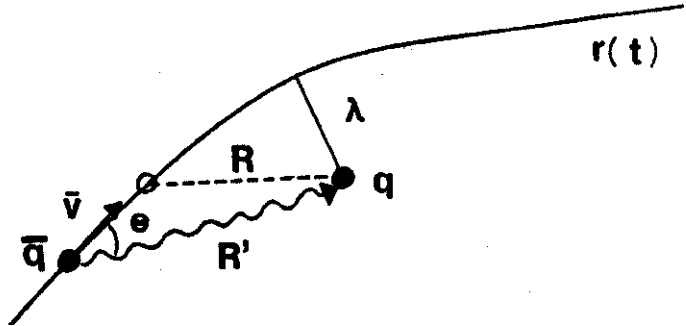

Figure. 1

$$R' \approx R(1 + \beta \cos\theta).$$

Let, the particles achieve the minimal distance $\lambda$ at time $t = 0$. The "effective" distance depends on the way the charges approach to each other:

$$R' \approx \lambda(1 + \beta \cos\theta). \qquad (8)$$

Denote by

$$t_{delay} = T\beta \cos\theta$$

the delay time along the direction of the dipole moment, varying between 0 and $T = \lambda/c$. $t_{delay}$ represents the time necessary for the signal to propagate along the dipole axis from the "effective" position $R'$ to the actual position $R$. From eq. (8), we get:

$$R' \approx R(t_{delay}) = \lambda\left(1 + \frac{t_{delay}}{T}\right) \qquad (9)$$

Following the DEMS model [1] and [3], the instant of minimal "effective" distance $\lambda$ between the charges is the instant of maximum luminosity of the source coinciding with the wavelength of the DEMS. Since the delay time is always within the interval $0 \leq t_{delay} < T$, from eq. (9) the ratio

$$\rho = \frac{R(t_{delay})}{\lambda} = 1 + \beta \cos\theta \qquad (10)$$

have values every close in the interval $1 \leq \rho < 2$, by using eq. (10), the eq. (5) modified becomes:

$$2E_s(E - (\rho - 1)mc^2) - \left(E^2 - (m\beta c^2)^2\right) = 0 \qquad (11)$$

Considering at very high energy a head-to-head collision, i.e. a collision occurring with an incoming incidence angle $\theta \cong 0$ and $\beta \cong 1$, eq. (10) converges to $\rho \cong 2$. In this case, DEMS is created with the maximum available energy $E_s = E$, where $E$ is the energy of the impinging particle 1 measured in the frame of the particle 2. Equation (11) becomes

$$(E - mc^2)^2 \cong 0 \qquad (12)$$

yielding for the impinging particle and for the total DEMS energy

$$E = mc^2. \qquad (13)$$

The result (13) means that for the observer in $S_2$ the maximum energy of the source is achieved when the relative velocity between the particles converges to speed of light $c$, i.e., $E_i = m_i c^2$ is the total energy carried by the particle 1 or symmetrically 2.

Since the measure of the total energy of the impinging particle is not the same for each inertial observer, in order to define a physical invariant we take in to account the non null numerator of the equation (6). Let

$$\varepsilon^2 = E^2 - P^2 c^2 \qquad (14)$$

be a non-null energy squared term due to non-impulsive energy contributions to the DEMS creation, so using eq. (13) and (14) we can define

$$E = \frac{\varepsilon}{\sqrt{1-\beta^2}} = mc^2 \qquad (15)$$

where $\varepsilon$ is the invariant energy of the impinging particle in its C.M, i.e. its rest energy.

Eqs. (14) proves that a DEMS can be energetically non-null only if the interacting particles have a non-null rest energy $\varepsilon$.

## 4) Relativistic Doppler Effect for interacting observers

A DEMSs distribution is produced only when a charged particle crossing matter interacts with anti-charged particles. During each collision, the DEMS produced allows the measurement of the dynamical state settled by the minimum distance $\lambda$ achieved between the interacting particles. Using eqs. (14) and (15), the measured photon energy (6), modified for an observer placed in the target frame, becomes

$$E_\gamma = \frac{1}{2} \varepsilon \frac{\sqrt{1-\beta^2}}{1-\beta\cos\theta} \qquad (16)$$

Considering that the energy (16) measured along the line of view of the observer 2 is that one of a characteristic DEMS produced during the interaction between the two observers, the characteristic wave frequency of the DEMS is dependent from the relative velocity $\beta$ between the interacting frames and from the interaction angle $\theta$ between the dipole axis coinciding with the line of view of the observer and the direction of motion of the moving frame. The frequency in the target frame of the observer is in agreement with one of a relativistic Doppler effect for a moving wave source:

$$\nu = \nu_0 \frac{\sqrt{1-\beta^2}}{1-\beta\cos\theta} \quad, \qquad (17)$$

where the rest frequency of the DEMS

$$\nu_0 = \frac{1}{2}\frac{\varepsilon}{h}$$

is half of the Compton's frequency of the impinging interacting particle.

Eq. (17) proves that:

(a) each particle feels an anti-particle as a wave;
(b) the rest frequency characterising the wave is half of the Compton frequency of the impinging particle.

**5) Relative observers and Lorentz's transformations**

Two observers perform a reciprocal measure of time and position $(x', y', z', t')$ respect the own frame, only if their frames are electromagnetically connected with a wave. To measure time and position, the distance must be greater or equal then a wavelength. If $\lambda$ and $T = \dfrac{\lambda}{c}$ are respectively wavelength and period of the source, the relative motion between the interacting frames implies a measure of period and wavelength modified to the Doppler effect. Using eq. (17) with $\nu_0 = 1/T$ and $\nu = 1/T'$, we obtain for two observers the fundamental measures of time and position inside the DEMS:

$$T' = \frac{T - \dfrac{\beta \cos\theta}{c}\lambda}{\sqrt{1-\beta^2}}$$

$$\lambda' = \frac{\lambda - \beta \cos\theta\, cT}{\sqrt{1-\beta^2}} \ ,$$

(18)

i.e. two observers cannot measure distances shorter than $\lambda'$. If two observers are located to a distance greater than the wavelength, i.e. the source is placed respectively at space-time coordinate $x' = a\lambda'$, $y'$, $z'$, $t' = aT'$ with $a > 1$, the wave front cover along the line of view jointing two observers a distance $x'$ with length within the interval $n\lambda' \leq x' < (n+1)\lambda'$. Multiplying eqs. (18) by the invariant factors

$$a = \frac{x'}{\lambda'} = \frac{x}{\lambda} = \frac{t'}{T'} = \frac{t}{T}$$

the measures of elapsed time from the signal and number of wave fronts between two observers, define symmetrically time and position of the moving frames that for great distances $1 \ll a < \infty$ gives the general Lorentz transformation

$$t' = \frac{t - \dfrac{\boldsymbol{\beta}\cdot\hat{\mathbf{r}}}{c}x}{\sqrt{1-\beta^2}}$$

$$x' = \frac{x \pm \boldsymbol{\beta}\cdot\hat{\mathbf{r}}\, ct}{\sqrt{1-\beta^2}}$$

(19)

When the relative motion occurs along the line of view of the observer, the angles are $\theta \cong 0$ or $\theta \cong 180°$, so standard Lorentz transformation defines symmetrically time and position of the two moving frames with a half wavelength precision:

$$t' = \frac{t \pm \frac{v}{c^2} x}{\sqrt{1-\beta^2}}$$

$$x' = \frac{x \pm v\,t}{\sqrt{1-\beta^2}}$$

(20)

The coordinates $y'$, $z'$ do not suffer changes because they are not involved in a relative motion along the line of view:

$$\begin{aligned} y' &= y \\ z' &= z \end{aligned}.$$

## 6) Conclusions

The derivation of eqs. (19) and the statement (I), prove that for a pair of observers associated with two charged particles in relative motion, the general Lorentz transformation connect their reciprocal space-time positions only assuming that each particle interacting with all the antiparticles of the pairs polarized in the vacuum medium. These interactions produce DEMSs, which waves, in agreement with eqs. (18) achieve the other observer with a period T' and a wavelength λ' contracted respect to ones emitted.

The effect symmetrically observed in each of the two inertial frames is analogous to one predicted by the relativistic Doppler effect, but springs out as a consequence of the energy and momentum conservation laws applied to the DEMS production during the transition $(p, \bar{p}) \rightarrow h\nu$ between the interacting particles and the exchanged photon.

From the point of view of two observers, each of the two consider itself at rest and the other in motion, so wavelength and period of the signals reciprocally exchanged define the metric of the space-time along the line of view connecting the two observers. This means that space-time structure derives from exchanged waves between all pair of points in which the DEMSs spontaneously produced during vacuum polarization are localised. Therefore, space-time with its metric and local curvature is created by waves emitted in the Ω phases of the DEMSs that have the role to connect all pairs of points defining an electromagnetic net in which each connection defines the local space-time metric.